\documentclass[12pt,fleqn]{elsarticle}
\usepackage{amssymb}
\usepackage{amsmath}
\usepackage{graphicx}
\usepackage{hyperref}
\usepackage{color}
\def\yen{{\setbox0=\hbox{Y}Y\kern-.97\wd0\vbox{\hrule height.1ex
width.98\wd0\kern.33ex\hrule height.1ex width.98\wd0\kern.45ex}}}
\begin{document}
\newcommand{\be}{\begin{equation}}
\newcommand{\ee}{\end{equation}}
\newcommand{\bc}{\begin{center}}
\newcommand{\ec}{\end{center}}
\newcommand{\bdm}{\begin{displaymath}}
\newcommand{\edm}{\end{displaymath}}
\newcommand{\ds}{\displaystyle}
\newcommand{\p}{\partial}
\newcommand{\INT}{\int\limits}
\newcommand{\SUM}{\sum\limits}
\newcommand{\bfm}[1]{\mbox{\boldmath $ #1 $}}
\renewcommand{\theequation}{\arabic{section}.\arabic{equation}}

\begin{frontmatter}

\author{G. Pontrelli\fnref{label1}\corref{cor7}}
\author{G. Toniolo \fnref{label2,label3}}
\author{S. McGinty\fnref{label4}}
\author{D. Peri\fnref{label1}} 
\author{S. Succi\fnref{label1,label5}}
\author{C. Chatgilialoglu\fnref{label2}}

\cortext[cor7]{Corresponding author: Email address: giuseppe.pontrelli@gmail.com}

\fntext[label1]{Istituto per le Applicazioni del Calcolo -- CNR
Via dei Taurini 19, Rome, Italy.}
\fntext[label2]{ISOF-CNR, Via P. Gobetti 101, Bologna, Italy.}
\fntext[label3]{Institute of Nanoscience and Nanotechnology, N.C.S.R. “Demokritos”, 15310 Agia Paraskevi Attikis, Greece.}
\fntext[label4]{Division of Biomedical Engineering, University of Glasgow, Glasgow, UK.}
\fntext[label5]{Italian Institute of Technology, CLNS@Sapienza, Roma, Italy.}

\title{\bf Mathematical modelling of drug delivery \\
from pH-responsive nanocontainers \normalfont }
		
\begin{abstract}
Drug delivery systems represent a promising strategy to treat cancer and to overcome
the side effects of chemotherapy. In particular, polymeric nanocontainers have attracted
major interest because of their structural and morphological advantages and the variety of polymers
that can be used, allowing the synthesis of materials capable of responding to
the biochemical alterations of the tumour microenvironment. 
While experimental methodologies can provide much insight, the generation of experimental data across a wide parameter space is usually prohibitively time consuming and/or expensive.  To better understand the influence of varying design parameters on the drug release profile and drug kinetics involved, appropriately-designed mathematical models are of great benefit. 
Here, we developed a novel mathematical model to describe drug transport within, and release from, a hollow nanocontainer consisting of a core and a pH-responsive polymeric shell. 
The two-layer mathematical model fully accounts for drug dissolution, diffusion and interaction with polymer. We generated experimental drug release profiles using daunorubicin and [Cu(TPMA)(Phenantroline)]$(ClO_4)_2$ as model drugs, for which the nanocontainers exhibited excellent encapsulation ability. The \textit{in vitro} drug release behaviour was studied under different 
conditions, where the system proved capable of responding to the selected pH stimuli by releasing a larger amount 
of drug in an acidic than in the physiological environments. By comparing the results of the mathematical model with our experimental data,  we were able to identify the model parameter values that best-fit the data and demonstrate that the model  is capable of describing the phenomena at hand.  The proposed methodology can be used to describe and predict the release profiles for a variety of drug delivery systems.

\end{abstract}


\end{frontmatter}
\section{Introduction}

The use of micro- and nano-particles as drug delivery systems is
an extensive area of research, but the full potential of such technology has yet to be realised.  There is growing interest in utilizing hydrogels, polymeric microspheres and nanoparticles
as carrier systems for cell-specific targeting and for `smart' delivery
of drugs, with potential advantages including the reduction in  systemic side-effects and an increase in drug efficacy [1-8].

Since their inception, nanoscale drug delivery systems (DDSs) have represented one of the most promising strategies to efficiently treat cancer and to overcome the unpleasant side-effects of conventional chemotherapy. The efficacy of cancer drugs is limited in clinical administration
due to their toxicity and poor solubility. Moreover,
intravenous injection and infusion are associated
with considerable fluctuation of drug concentration in
the blood. Therefore, the drugs can only be administered
over a limited dosage and time period. The intrinsic ability of DDSs to target tumor tissues and cells relies on the so-called enhanced penetration and retention (EPR) effect \cite{mae}. This is the underlying basis for the employment of DDSs as amelioration for cancer therapy, since it takes advantage of the unique features of tumor tissues to  direct the drug to its target. Specifically, DDSs targeting cancer should accumulate in tumours precisely because of the pathophysiological differences between tumor and healthy tissues. Due to overstimulated and defective angiogenesis, tumours have leaky vessels allowing the penetration of DDSs with compatible size \cite{all}. Also, the inefficient lymphatic drainage guarantees the retention of the delivery agents and the released drug in the area \cite{fang,bran}. Whereas more conventional drug administration is indiscriminate, nanoscale DDSs are designed to specifically enter and accumulate into the tumor tissues because of the EPR effect, where they are meant to release their payload. \par

The effectiveness of nanoscale polymeric delivery systems can be improved by designing structures capable of responding to specific pre-set conditions by altering their properties and favoring the release of the loaded drug.
Stimuli-responsive nanocontainers are a family of DDSs
that can control the release of the therapeutic active agents in response
to external triggers such as temperature, pH, electrical fields and many others. In particular,
when the pore geometry or chemical composition can be altered
in response to environmental stimuli, the nanocontainer is
expected to provide a wide range of applications because of the
selective permeability and controlled release of its cargo \cite{mat} . Therefore, the development
of new kinds of environmental stimuli-responsive and “smart” 
drug delivery systems is relevant and highly desirable. They have gained increasing attention 
recently and many examples can be found in literature \cite{yan,zha,ton1,ton2,song}. 

Tumor cells and tissues are characterized by some internal biochemical alterations 
that can be used as a trigger for drug release \cite{flei,gan}. Among these alterations, the best known and most 
exploited one is probably pH. There is a clear difference between healthy tissues (pH $\approx$ 7.4) and diseased tissues (pH $<6.0$ in tumours). 
Also, intracellular differences between normal and cancer cells have been highlighted and 
can be used to facilitate drug release \cite{yan, zhan}. 
\begin{figure}[ht!]
\label{fig1}
\centering
\includegraphics[scale=0.6]{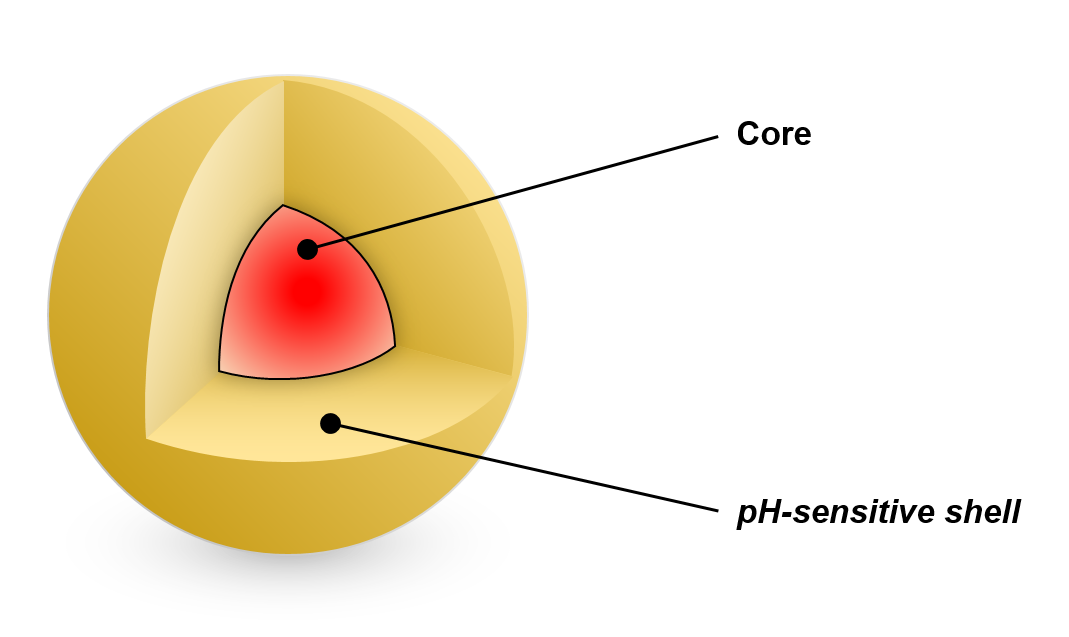} 
\caption{3D representation of a core-shell nanocontainer (figure not to scale).}
\end{figure}
At a tissue level, tumours have lower extracellular pH due to their faster metabolism and lower oxygen content. 
Lack of oxygen may cause hypoxia, leading to the production of lactic acid, which in turn reduces pH in the tissue \cite{gan}. 
At a cellular level, in the endosome-lysosome system, the preferential route for the internalization, 
DDSs encounter much lower pH than the neutral extracellular environment, such as the endosome (pH 5.0--6.0) and lysosomes (pH 4.5--5.0). 
Lysosomal pH in cancer has been reported to be as low as 4.0 
 \cite{eft}.  \par
Polymers that display a physico-chemical response to stimuli have been
widely explored as potential DDSs \cite{wang}. 
For example, ionizable polymers are optimal candidates for the synthesis of pH-responsive systems. 
Weak acids and bases such as carboxylic acids and amines exhibit a change in the ionization state depending on variation of the pH, which in turn affects the capability of the DDS to interact with drugs. Therefore, pH variations can result in breakage of these interactions, facilitating the release of a loaded drug \cite{sch}. Among a variety of DDSs, one of the most 
promising examples are core-shell nanocontainers (NCs), due to their ability to carry larger amount of drugs than other systems \cite{song} . Typically, a core-shell NC consists of a drug-loaded (fluid or solid) spherical centre ({\em core}) coated by a polymeric layer ({\em shell}) acting as a protective barrier against external chemical aggression and mechanical erosion. 
The core structure is generally conceived to locate the therapeutic agent, 
whilst the polymer shell is designed to control the drug release (Fig. 1). The drug is encapsulated in both compartments but the core is known to be extremely important to increase the amount of loaded drug compared to other systems \cite{kan}. Such two-layer assembly allows for better control of the drug release. 

Mathematical and computational (\textit{in silico}) modelling can provide a better understanding of the influence of different design parameters,
which may then either be used to reduce the design space for experiments or, more ambitiously, be utilised as a predictive screening tool for drug carriers \cite{wang,siepmann3,fredenberg}.
\begin{figure}[ht!]
\label{fig2}
\centering
\includegraphics[height=0.27\textwidth]{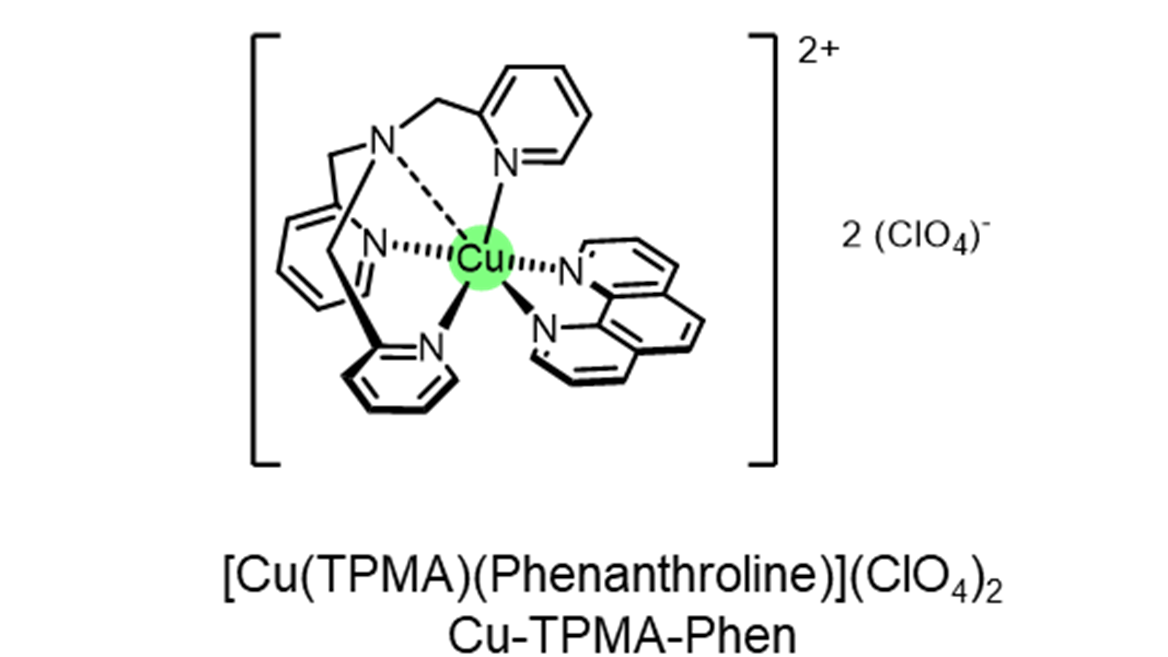}
\hspace{0.0cm}\includegraphics[height=0.27\textwidth]{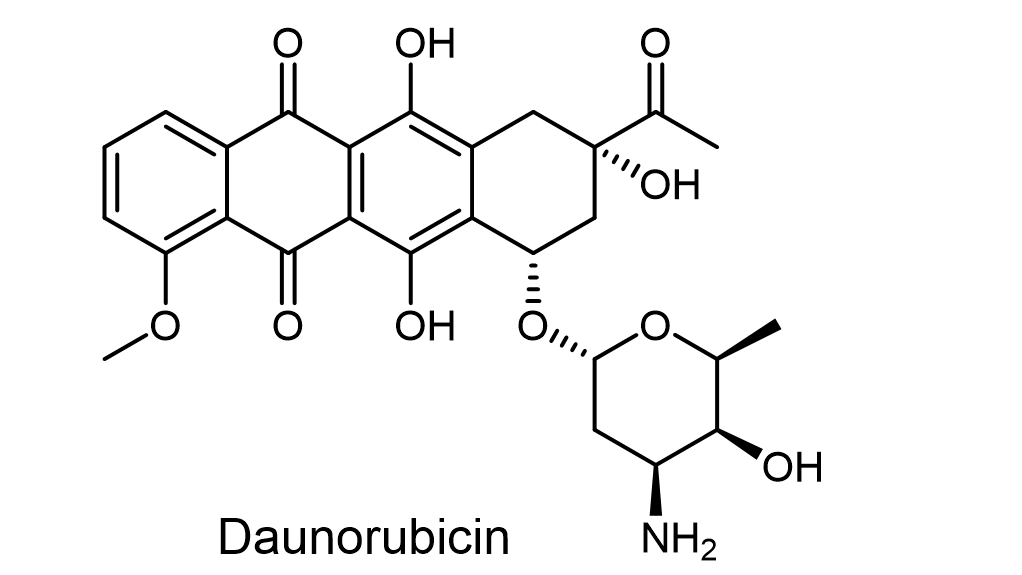}
\caption{Chemical structure of CTP  (left) and DNR (right).}
\end{figure}
 Numerous mathematical models have been derived to describe
and predict drug release from  diffusion controlled-release systems, accounting for the various physical and chemical
phenomena that contribute to the overall drug release kinetics \cite{xu,wang,siepmann3}.  However,  theoretical studies on drug
delivery from pH-responsive systems are relatively scarce.  One exception is the study by Manga et al. \cite{man}, who considered the effect of pH on drug release from hydrogels  by modelling a pH-dependent swelling behaviour.

In this paper, we develop a novel mathematical model of drug transport within, and release from, a drug-loaded NC.  The model considers the two distinct layers (core and shell) and accounts for drug dissolution and diffusion within the core, as well as drug diffusion and binding within the shell. Several of the models parameters are considered pH-dependent, enabling the model to account for pH-dependent drug release.  The model is calibrated through comparison of the model results with existing and new experimental drug release data from pH-sensitive NCs containing two different molecules (Fig. 2): (i) [Cu(TPMA)(Phenantroline)]$(ClO_4)_2$ (hereafter referred
to as CTP), a highly innovative metallodrug recently documented for gene therapy \cite{fan}  (drug release data reported in  \cite{ton2}) and (ii) the chemotherapeutic agent daunorubicin (DNR) (new drug release data presented here). While the synthesis of the NC's has already been reported, the present study focuses on the release kinetics of these two drugs from the NCs, since DNR is currently one of the most used chemotherapy agents while CTP is a very promising candidate for future generation medicine which would greatly benefit from selective release in the target area [15].

\section{Materials and methods}
\setcounter{equation}{0}
\subsection{Synthesis of pH-responsive nanocontainers}
The synthesis of the pH-responsive nanocontainers has already been reported \cite{ton2}. Here we provide a brief summary and point the reader to the supplementary data for further details.  We consider drug release data from a pH-sensitive hollow NC, predominantly made of  methacrylic acid (MAA), the monomer responsible for the pH sensitivity. The carboxylic groups of the resulting poly methacrylic acid (PMAA) shell can be protonated or deprotonated depending on the pH, enabling different interactions with the external environment
\cite{qiu, zhan2}. To form the shell, we utilised two  additional monomers: N,N-methylenebis(acrylamide) (MBA), used as a cross-linking agent to maintain the structure of the hollow NCs in water, and poly(ethylene glycol) methyl ether methacrylate (PEGMA), which is a hydrophilic, nontoxic component known to show resistance against nonspecific protein adsorption and to prolong the \textit{in vivo} residence time of the DDS \cite{bila}. 

The resulting diameters of the central cavity and of the shell suspended in water were 0.3$\mu$m and 0.55$\mu$m, respectively, according to the requirements for such systems. A relatively constant diameter after long-term storage at room temperature was observed,
indicating favorable stability properties. No effects of erosion or degradation were reported over the time scale considered.

\subsection{Drug loading}
In the present study we loaded the NCs with DNR, while in a previous study our model drug was CTP \cite{ton2}.  For DNR loading, 5mg of hollow NCs were suspended in 5ml of phosphate buffer saline (PBS, pH 7.4) with the aid of ultrasonic bathing. 5mg of daunorubicin hydrochloride (DNR HCl) were then dissolved in the medium. The suspension was covered with foil and maintained under gentle agitation for 72h at r.t. The non-encapsulated DNR was then removed with 15 cycles of centrifugation/resuspension (5min $\times$ 9000rpm). The encapsulated amount of DNR was indirectly determined by UV spectroscopy: the total amount of loaded DNR was calculated by the difference between the amount of DNR in feeding and in the supernatant fractions. These calculations were based on a standard curve of DNR in PBS and the concentration was determined with absorbance measurements at 484nm.  The drug loading process was evaluated using the parameters encapsulation efficiency (EE\%) and loading capacity (LC\%), defined as:

\begin{center}
\noindent EE\%=Encapsulated drug (mg) / drug in feeding (mg) $\times$ 100  
\medskip \\ 
\qquad LC\%=Encapsulated drug (mg) /drug loaded into NCs (mg) $\times$ 100 
\end{center}

\noindent For DNR, we obtained EE(\%)$=87.1 \pm 2.9$ and LC(\%)$=47.2 \pm 0.1$, which translates into an encapsulation of 0.893mg of DNR per 1mg of NC. The loading ability of the hollow NCs relies on the interactions between the groups of the shell and the drug. Specifically, electrostatic interaction is the most important one and involves the negatively charged carboxylic groups of PMAA (pKa ca. 4.5; in the anionic form $-COO^{-}$ in the loading conditions) and the amino groups of the drug (pKa 8.6 \cite{kira}; mostly as $- NH^{3+}$ in the loading conditions). Hydrogen bonds also play a crucial role, which involve other non-ionized functional groups, such as the amide group of the cross-linking agent, PEG chains and the carbonyl and hydroxyl groups of 
DNR \cite{meta}. The encapsulation process of CTP (see supplementary material) resulted in encapsulation efficiency and loading capacity  of EE(\%)$=36.4 \pm 5.2$ and LC(\%)$=42.0 \pm 3.3$, respectively, which corresponds to the encapsulation of 0.724 mg ($0.988 \mu$mol) of CTP  per 1mg of NC \cite{ton2}. Unlike DNR, CTP is water-insoluble and 
 adequate aqueous conditions to dissolve the drug and to suspend the NCs were required. This was achieved by adding 5\% of acetronitile to the loading buffer (PBS). The main interaction involved in the encapsulation of CTP is probably electrostatic between the negatively charged carboxylate anions of PMAA and the 2+ positive charge of Cu, which the complex assumes upon dissolution of the perchlorate counterions. In addition, in view of previously-reported data on this molecule, a distal pyridine nitrogen donor atom of TMPA was identified within the coordination complex and may therefore interact with the NCs through hydrogen bonding \cite{fan}. \par

\subsection{\textit{In vitro} drug release studies}
The pH sensitivity of the system was tested by means of  {\em in vitro} drug release experiments. For the DNR experiments, 1mg of DNR-loaded NCs were suspended in 0.5ml of buffer and loaded into MWCO 140 kDa dialysis tube and incubated in 50ml each buffer solution. 
Three different pH conditions were used for the analysis of release: citrate buffer 0.1M pH 4.0 -- 5.5 and PBS 1$\times$ pH 7.4. At different time points (30min, 1h, 2h, 5h, 8h, 10h, 24h, 48h, 72h), 1ml of the solutions was withdrawn and analyzed. The concentration of each sample, and therefore of the release medium at each time point was determined with UV spectroscopy by using the standard curve method (at 484 nm). A standard curve of DNR was recorded in each buffer used as a release medium. The experiment was carried out three times for statistical analysis. For details of the CTP drug release study that we previously reported \cite{ton2}, we refer the reader to the supplementary material.

\section {Mathematical modelling}
\setcounter{equation}{0}
\subsection{Modelling drug release from core-shell nanocontainers}
Our previous study on CTP release from the NCs revealed that the amount of drug released varies depending on the pH of the environment. The CTP release profile from the NCs was studied in both acidic and slightly basic environments. After 24h, the amount of  drug released was, respectively, 50\% and 32\% \cite{ton2}. The carboxylic groups of PMAA are the key for the interpretation of these results. They are mostly protonated at pH 4.0 and they cannot interact with the positively charged CTP complex, causing the release. In addition, it is known that at pH 4.0, the hydrogen-bonding interactions are weaker than in neutral conditions, thus facilitating the release \cite{yan}. 
Motivated by these findings, here we develop a reaction-diffusion continuum model to describe drug (either CTP or DNR) transport within, and release from, the drug-loaded NCs under different pH conditions. We restrict our attention to modelling \textit{in vitro} drug release so that we may effect comparison with our experimental data.  We note that no significant degradation of the NCs was observed over the course of the \textit{in vitro} experiments. 

We start by considering  a single NC as a two-layer spherical system, comprising an internal core $\Omega_0$ and the enveloping concentric shell $\Omega_1$. Let us denote by $R_0$ and $R_1$ the internal and external radius of the NC, with the origin located at the centre of the NC and the $r$-axis oriented with the positive direction pointing outwards (Fig. 3).  In what follows, the subscripts $0$ and $1$ indicate parameters and variables referring to the core and shell layer, respectively.
Due to the homogeneity and isotropy of each layer,  we can assume that net drug diffusion occurs along the radial direction only, 
and thus we restrict our study to a one-dimensional model that reflects a perfectly radially symmetric system. \par

\begin{figure}[ht]
\label{fig4}
\centering
\includegraphics[scale=1]{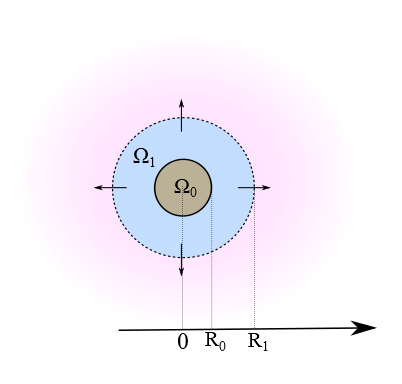} 
\caption{Schematic representation of a cross-section of the two-layer NC, comprising an internal core $\Omega_0$
and an external shell $\Omega_1$ (figure not to scale).}
\end{figure}

The majority of the drug is contained within the core, and we assume an initially homogeneous distribution of drug within this region, at some concentration $B_0$.  However, the particle preparation methods may also result in some drug mass being contained with the shell initially. We assume that this drug, of concentration $B_1$, is permanently encapsulated and may never be released. When exposed to the release medium, the NC uptakes water and a dissolution process ensues, converting immobile (undissolved) drug of concentration $b_0(r,t)$ in the core to dissolved drug of concentration $c_0(r,t)$.  Following our previous work \cite{pon1}, we model dissolution as a nonlinear process whereby drug dissolves at a rate $\beta$ and in proportion to the difference between the dissolved drug concentration and the  solubility of the drug in the medium, $S$. The $2/3$ exponent accounts for potential influences on the dissolution rate as the surface area of the dissolving drug particles changes \cite{frenning}.  When dissolved,  the drug is able to diffuse through the core with diffusion coefficient $D_0$.
The dynamics of drug dissolution and diffusion in the core is then described by the following two partial differential equations:
\begin{align}
& {\p b_0\over \p t} = -\beta b_0^{2/3}\left(S-c_0\right), \nonumber  \\
 &{\p c_0 \over \p t} = D_0 \left( {\p^2 c_0 \over \p r^2} +    {2 \over r}  {\p c_0 \over \p r} \right) 
+ \beta b_0^{2/3}\left(S-c_0 \right),    \qquad \mbox{in}\, \, (0, R_0).  \label{eq22}
\end{align}
 
\noindent Our experiments clearly show that a pH-dependent fraction of the initial drug loading is retained within the NCs and is never released.  We model this observed phenomenon through first order reaction kinetics, whereby drug diffusing through the polymeric shell has the possibility to permanently bind to the polymeric shell at a rate $k$.  We note that other forms of interaction could have been considered: in this study, however, we chose to focus on a simple linear reaction model in the absence of evidence to suggest otherwise. Drug that is able to traverse the full  
 radius of the shell is rapidly cleared due to the sink conditions imposed in the experiment.  Denoting by $b_1(r,t)$ and $c_1(r,t)$ the bound and unbound phase concentrations, respectively, the drug dynamics in the shell are then governed by the following equations \cite{pon1}:
\begin{align}
&{\p c_1 \over \p t} = D_1 \left( \ds{\p^2 c_1 \over \p r^2} +   {2 \over r}  {\p c_1 \over \p r} \right) - k c_1  \nonumber \\
&{\p b_1 \over \p t} =k c_1  \qquad \mbox{in}  \,\, (R_0, R_1), \label{eq44}
\end{align}
\noindent where $D_1$  represents the diffusion coefficient of the drug in the shell.
To close the  system (\ref{eq22})--(\ref{eq44}),  we are required to impose appropriate boundary and initial conditions.  At the interface between the core and shell layers, we assume continuity of flux and concentration:
\be
  -D_0{ \p c_0 \over \p r} = -D_{1}{ \p c_{1} \over \p r},  \qquad\qquad  c_0=c_1  \qquad \mbox{at} \, \, r=R_0.  \label{eru34}
 \ee 
For radial symmetry we require:
\be
 {\p c_0 \over \p r} =0  \qquad\qquad\qquad\quad \mbox{at $r=0$.}  \label{ad1}
\ee
At the NC surface, we impose a perfect sink condition because of the relatively large size of the release medium and well stirred conditions:
\be
c_1=0, \qquad\qquad\qquad\quad \mbox{at $r=R_1$,}  \label{ad2}
\ee
At initial time, the drug is loaded in the core at concentration $B_0$, while the shell contains bound drug at concentration $B_1$:
\be
b_0=B_0, \qquad \quad c_0=0    \qquad \quad   b_1=B_1 \qquad \quad c_1=0   \label{prob20}
\ee

The total mass of drug within the NC at any time is given by integrating the concentration of drug within each phase and layer  over the corresponding volume \cite{pon1}, that is 
\be
M_{tot}(t)=4\pi\left[\int\limits_{0}^{R_0} r^2 \left\{b_0 (r,t)+c_0(r,t)\right\} dr  +\int\limits_{R_0}^{R_1} r^2 \left\{b_1 (r,t)+c_1(r,t)\right\} dr\right]
\ee
The drug release profile, $M_{rel}(t)$, defined as the cumulative $\%$ of drug released by time $t$, is then given by
\be
\%M_{rel }(t)= {M_{tot}(0) - M_{tot}(t) \over  M_{tot}(0)} \, \times \, 100,
\label{releaseprofile}
\ee
 where $M_{tot}(0)$ is the total initial mass of drug in the NC.

\subsection{Solution method}
Before solving the model (\ref{eq22})--(\ref{prob20}) numerically, it is convenient first to nondimensionalise the equations. We scale $r$ with the radius of the shell and scale $t$ with the timescale for diffusion in the shell: 
\be
r \rightarrow {r \over R_1},   \qquad t \rightarrow {D_1 \over R_1^2} t.
\ee

\noindent Scaling all concentrations with $B_0$, the model may then be written in terms of five non-dimensional groups:
\be
 D=\frac{D_0}{D_1},   \qquad Da=\frac{\beta B_0^{2/3}R_1^2}{D_1},   \qquad   \tilde k =  {k R_1^2 \over D_1},   \qquad   
\tilde S = {S \over B_0},  \qquad  \tilde B_1={B_1 \over B_0},  \label{param1}
\ee
where $Da$ may be regarded as a Damk\"{o}hler number, defined as the ratio of dissolution rate to diffusion rate,
and  $\tilde k$ denotes the ratio between the binding rate and rate of diffusion in the shell. 
The  nondimensional system of six equations were then discretized spatially before solving the resulting system of ordinary differential equations, following the method we previously described \cite{pon1} (for further details please refer to the supplementary material).

\subsection{Optimization strategy}
As in many biological systems, the model
contains a number of parameters: many of these are not known \textit{a priori}, and those that have been measured are often subject to high variability and uncertainty. Obtaining reliable estimates of parameters is a significant challenge in the field.  Starting from a physically realistic range of parameters (Table 1), we address this issue here by an optimization strategy.  Specifically, 
using the experimental data points, we inversely estimate the five unknown nondimensional parameters (\ref{param1}) of the model for each pH,
such that the model solution best fits the data for each drug.  We employ an optimization method based on an objective function subject to a number of constraints (due to physical and experimental conditions) over a set of experimental data.  We proceed first with a direct search  based on a standard least squares approach and then refine
 with a pattern-search algorithm. For further details, we refer the reader to the supplementary material.

\begin{table}
\label{interv1}
\begin{center}
\caption{Possible range of the nondimensional parameters considered in optimization algorithm. These ranges were chosen based on physical constraints and typical values, and span at least 3 orders of magnitude for each parameter.}
\label {opt0}
\begin{tabular}{|l|l|l|l|} \hline\hline
          Parameter    &  Min.               & Max.               \\ \hline\hline
$D$    & $2$ & $10^{3}$ \\ \hline\hline
$Da$       &  $10^{-1}$ & $10^{2}$ \\ \hline
$\tilde S$          &  $10^{-2}$ & $10^{2}$ \\ \hline
$\tilde k$          &  $10^{-2}$ & $10^{2}$ \\ \hline
$\tilde B_1$ &  $10^{-3}$ & $10$ \\ \hline
\end{tabular}
\end{center}
\end{table}

\section{Results and discussion}
\label{numerical}
\setcounter{equation}{0}
\subsection{DNR \textit{In vitro} drug release}
The results of our \textit{in vitro} DNR release study clearly show, similarly to CTP, that the amount of drug released varies depending on the pH of the environment. This behavior can be explained by considering the interaction between DNR and the carboxylic groups of the NCs. At physiological pH (7.4) the carboxylic groups of PMAA are deprotonated and interact with the protonated amino group of DNR (pKa 8.6) favoring the retention of the drug in the NCs (also encapsulation conditions). At pH 5.5, a lower percentage of carboxylic groups are protonated, which causes a decrease in the number of interactions DDS-DNR and, as a consequence, a larger amount of drug is released. At pH 4.0 the majority of the carboxylic groups are  protonated, therefore there will be fewer electrostatic interactions and the drug will be more easily released than in the two previously-described conditions.

\subsection{Drug release simulation}
The results of our parameter identification procedure are detailed in Table \ref{opt1} and Table \ref{opt2} for the drugs DNR and CTP, respectively.  For DNR we observe an increase in the value of four of the non-dimensional parameters ($D$, $Da$, $\tilde{S}$ and $\tilde{k}$) with increasing pH, over the values of pH studied in the experiments (Fig. 4).  
 The value of $\tilde{B_1}$, however, remains constant since this is simply the ratio of initial drug concentration in the shell normalised by the initial drug concentration in the core and is a function of the fabrication process rather than the pH.  Within the supplementary material, we explain in detail how this parameter is obtained as part of the inverse problem.  Similar trends are observed for CTP for the parameters $D$, $Da$ and $\tilde{k}$ (Fig. 4).  However, for CTP, the normalised solubility $\tilde{S}$ decreases with increasing pH, according to what expected due to the complete protonation of amine and pyridine nitrogens at low pH.
 An interesting result from Tables \ref{opt1}-\ref{opt2} is the monotonicity of the parameters with pH.  The order of magnitude of the parameters for DNR and CTP is the same: due to the extremely large space of parameters, this is an indirect confirmation of the correctness of the optimization procedure.

\begin{table}
\label{tab1}
\begin{center}
\caption{Optimal nondimensional parameters at three values of pH for DNR.}
\label{opt1}
\begin{tabular}{|l|l|c|c|c|} \hline\hline
   & Parameter  & pH=4             & pH=5.5     & pH=7.4           \\ \hline\hline
1  & $D$       & $35.60$  & $61.50$  & $113.05$ \\ \hline\hline
2  & $Da $   & $0.63$  & $0.96$ & $1.10$  \\ \hline
3  & $\tilde S$ & $2.64$ & $3.35$ & $5.60$ \\ \hline
4  & $\tilde k$  & $0.47$  & $4.65$  & $10.89$ \\ \hline
5  & $\tilde B_1$ & $0.12$  & $0.12$ & $0.12$  \\ \hline
\end{tabular}
\end{center}
\end{table}

\begin{table}
\label{tab2}
\begin{center}
\caption{Optimal nondimensional parameters at two values of pH for CTP.}
\label{opt2}
\begin{tabular}{|l|l|c|c|} \hline\hline
   & Parameter  & pH=4        & pH=7.4           \\ \hline\hline
1  & $D$       & $48.59$   & $86.91$ \\ \hline\hline
2  & $Da $   & $0.14$  & $0.67$  \\ \hline
3  & $\tilde S$ & $2.20$  & $0.11$ \\ \hline
4  & $\tilde k$  & $0.09$  & $6.31$ \\ \hline
5  & $\tilde B_1$ & $0.18$  & $0.18$  \\ \hline
\end{tabular}
\end{center}
\end{table}

\begin{figure}[ht]
\centering
{\includegraphics[scale=0.2]{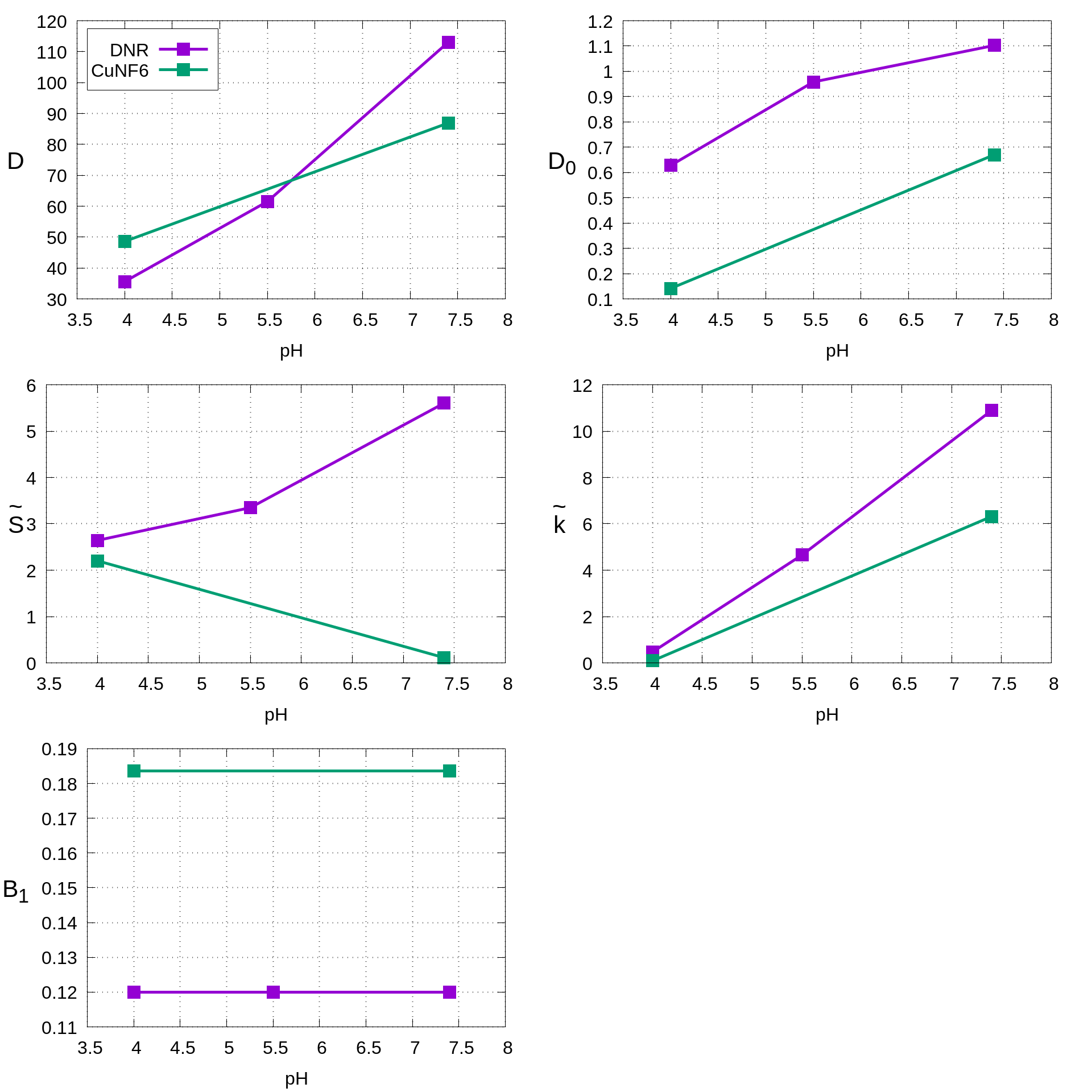}}
\caption{Variation of the five nondimensional parameters vs pH.}
\label{fig7}
\end{figure}

\begin{figure}[ht!]
\centering
{\includegraphics[scale=0.23]{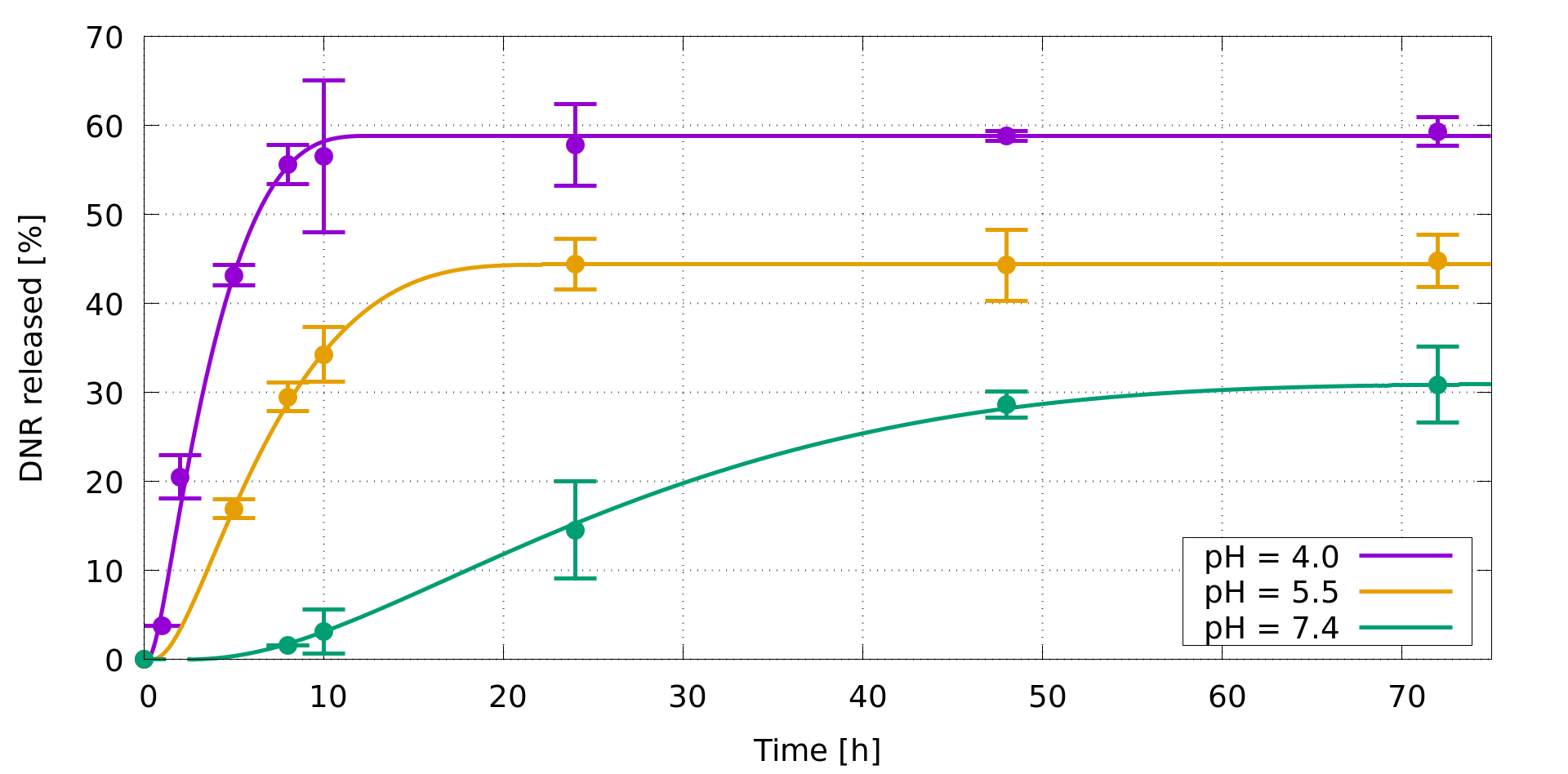}}
\caption{Best fitting release curves vs. experimental data for DNR at pH=4, 5.5 and 7.4. }
\label{fig5}
\end{figure}

\begin{figure}[ht!]
\centering
{\includegraphics[scale=0.23]{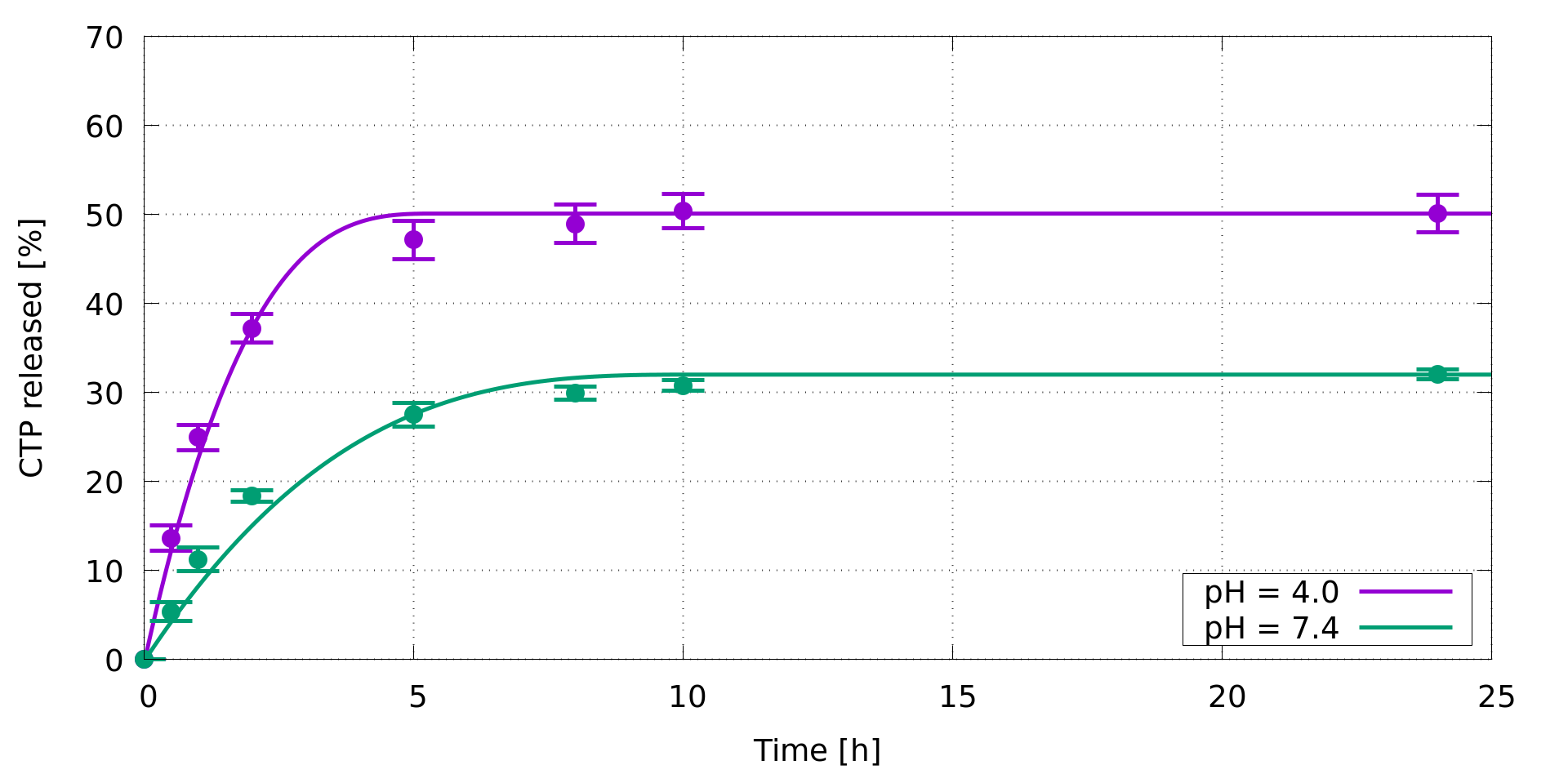}}
\caption{Best fitting release curves vs. experimental data for CTP at pH=4 and 7.4. (experimental data
taken from \cite{ton2}). }
\label{fig6}
\end{figure}

In Figures \ref{fig5}-\ref{fig6} we show the experimental drug release data for DNR and CTP.  The curves correspond to the solution of the model obtained with the optimal parameters for the values of pH studied.  Clearly, the release of each drug is well-captured by the two-phase two-layer dissolution-diffusion-reaction model that we have devised. Probing further, we are able to establish that the slower release of DNR with increasing pH is likely as a result of a slower diffusion coefficient in the shell, coupled with faster binding to components of the shell.  As a result, as the pH is increased, a greater fraction of the initial drug load is permanently retained and never released.  For CTP, the picture is a little more complicated. Firstly, the decrease in solubility with pH has the effect of slowing the dissolution process.  However, there is a modest increase in diffusion coefficient within the shell with pH, which coupled with the the simultaneous increase in binding within the shell results in a $\tilde{k}$ that greatly exceeds 1, indicating that binding is dominating and transport within  the shell is increasingly diffusion-limited. \par

Our mechanistic model confirms a pH-responsiveness of the PMAA shell in a manner that is dependent on the particular drug studied. For DNR we observe an increasingly pronounced delay in drug-release with pH, likely corresponding to the slower diffusion coefficient and faster binding in the shell with pH, as described above. This effect is significantly less for CTP, where we observe an ``initial burst'' of drug, particularly for the lowest value of pH,  which may be beneficial if an immediate and fast deliver of drug is desired, as in presence of cancer cells. 
The implication is that, while for targeting cancer the biological effect of the drug is important, the release kinetics of the drugs can vary depending on their chemical-physical properties and interaction with polymer, meaning that both aspects have to be considered hand-in-hand when choosing an appropriate drug to load the NC.

\subsection{Limitations}
We would like to emphasize that there are limitations in this work.  The mathematical model makes a number of assumptions as detailed in the text and the experimental data have been obtained in an $\textit{in vitro}$ environment. Importantly, while we have demonstrated that a dissolution-diffusion-reaction mechanism captures the drug release well, the different identified model parameters for different values of pH points to a complex relationship between pH and the various drug-transport parameters.  In this preliminary study, we have not sought to identify the particular functional dependence of the various parameters on pH, and this is left for future work. Notwithstanding, the approach adopted here of identifying these parameters computationally on a small set of \textit{in vitro} data is still very useful, since once calibrated the model can be used in a predictive sense to reduce the number of \textit{in vitro} experiments, and with further modification, can be correlated with $\textit{in vivo}$ data.

\section{Conclusions}
Nanocontainers made of pH-responsive polymers are of technological interest and have many
potential applications in medicine, including their use as controlled DDSs. They have attracted 
great attention as potential drug
carriers and offer significant advantages over more traditional therapies, 
both in terms of efficacy and safety.
We have shown that a pH sensitive controlled-drug releasing NC is  
well described by a dissolution-diffusion-reaction core-shell
mathematical model. 

 A number of release  experiments involving dispersed NCs of identical size and initial loaded drug have been
carried out and the data have been used to inversely estimate the best-fitting parameters of the model for each pH studied. 
The different physico-chemical characteristics of the two drugs
affect their interactions with the pH-sensitive NCs that in turn, influences the release performance.
This is reflected through the parameters of the mathematical model. 
Once these parameters have been computationally identified, the proposed methodology
provides a simple tool that can be used to quantitatively characterize
the drug kinetics, improve the technological performance and
optimize the release rate for the target application. Our combined \textit{ in vitro} experimental and mathematical modelling framework may be
used to predict NC drug release performance, opening up the possibility of an \textit{in silic}o approach to optimising the drug
release profile and ultimately the effectiveness of these devices. 


 
\vspace{1cm} 
\noindent{\bf Acknowledgments}  \\
Funding from the European Research Council under the European Unions Horizon
2020 Framework Programme (No. FP/2014-2020)/ERC Grant Agreement No. 739964 (COPMAT) is acknowledged.  \\
C.C. and G.T. acknowledge funding from the Marie Skłodowska-Curie
Innovative Training Network (ITN) ClickGene
(H2020-MSCA-ITN-2014-642023). 
We are grateful to Drs. E.K. Efthimiadou and G. Kordas for helpful
discussions regarding section 2.
\vspace{1cm} 

\newpage
\appendix
\section{Supplementary data}
\setcounter{equation}{0}
\renewcommand{\theequation}{A.\arabic{equation}}
\subsection{Numerical discretization}

The nondimensional model is given by:
\begin{align}
& {\p b_0\over \p t} = - Da \, b_0^{2/3}\left(\tilde S-c_0\right) \qquad &\mbox{in}\, \, (0, R_0), \label{er21}   \\
 &{\p c_0 \over \p t} = D \left( {\p^2 c_0 \over \p r^2} +    {2 \over r}  {\p c_0 \over \p r} \right) 
+ Da \, b_0^{2/3}\left(\tilde S-c_0 \right),    \qquad &\mbox{in}\, \, (0, R_0),  \label{er22} \\
&{\p c_1 \over \p t} =  {\p^2 c_1 \over \p r^2} +   {2 \over r}  {\p c_1 \over \p r}  - \tilde k c_1  \qquad &\mbox{in}  \,\, (R_0, 1),  \label{er43} \\
&{\p b_1 \over \p t} = \tilde k c_1  \qquad &\mbox{in }  \,\, (R_0, 1),   \label{er44} \\
&{\p c_0 \over \p r} =0  \qquad &\mbox{at  } r=0,   \label{ad100}  \\
& -D { \p c_0 \over \p r} = -{ \p c_{1} \over \p r},  \qquad\qquad  c_0=c_1  \qquad &\mbox{at  } r=R_0,  \label{er45}  \\
&c_1=0, \qquad &\mbox{at  } r=1.  \label{ad200}
\end{align}

We proceed to solve the system of equations (\ref{er21})-(\ref{ad200})  numerically, building on the method we described previously \cite{pon1}.
Let us subdivide the interval $(0,R_0)$ into  $N+1$ equispaced
grid nodes,  and the interval $(R_0,1)$ into $M+1$ equispaced  points, with $h_0$ and $h_1$  
the spacing in the core and shell layers, respectively.  
Let us indicate by a superscript $j$ the approximated value of the concentrations  at $r_j$. 
In each layer, we approximate the diffusive terms by considering a standard second order central difference in space of the second derivative at internal nodes.
The reaction terms in eqn. (\ref{er21})-(\ref{er44})  do not contain any spatial derivatives and therefore are evaluated pointwise. 
For example,  (\ref{er22}) is discretized at node $r_j$ as:
\be
\left.{dc_0 \over dt}\right |_{r_j} =  D {c_0^{j-1} - 2 c_0^{j} +c_0^{j+1} \over h_0^2}
 + Da \, (b_0^j)^{2/3} (\tilde S-c_0^j).
\ee

\noindent After spatial discretization, the system of PDEs reduces to a set of nonlinear ordinary differential eqns. of the form:
\be
{dY \over dt} = A ( Y)  \label{ew1},
\ee
where $Y=(b_0^0 ,....   ,b_0^{N-1},c_0^0 ,....   ,c_0^{N-1} , c_1^1 , ...   ,c_1^M , b_1^1, ....   b_1^M )^T$
and $A(Y)$ contains the discretized eqns. (\ref{er21})-(\ref{er44})     and related boundary/interface conditions (\ref{ad100})-(\ref{ad200}) .
The system (\ref{ew1}) is solved by the routine ode15s of {\tt Matlab} based on a Runge-Kutta type method with backward differentiation formulas, 
and an adaptive time step \cite{pon1}.

\newpage
\subsection{Parameter identification }
\label{optimization}

The mathematical model described in Section 3 and its successful use relies on the knowledge of the physico-chemical parameters. Unfortunately, the determination of these quantities by means of experiments is impractical and often associated with large error.
As a consequence, we derive the unknown parameters by solving an inverse problem: the experimental data are compared with the prediction of numerical model, and the parameters are used as independent variables to minimize the distance between the experimental data and the numerical prediction. An optimization problem is then formulated: given  $N_{s}$ experimental samples $X_i=X(t_i)$ (drug released) measured at different times $t_i$, we define our objective function by a least square method:
\[
      F(\xi) = \sum_{i=1}^{N_{s}} (X_i-x_i(\xi))^2
\]
where $x_i(\xi)$ correspond to the computed quantities, depending on the unknown parameter set $\xi$.
Then, we can minimize $F$  subject to a number of constraints and $\xi$ in a given range.
Due to the large range investigated, some combinations of the parameters could be physically unrealistic, and consequently produce unphysical results that should be discarded. To address this, two different constraint functions are adopted here, the first related to the released mass, that cannot be negative, and the second enforced on the positiveness of the first derivative of the drug release curve, since a negative first derivative would imply that released drug re-enters the NC.
Unfortunately, the space of the parameters $\xi$ is very large: this poses some further difficulties in the optimization problem, increasing also the number of areas where the design parameters produce good values of $F(\xi)$. Taking account of the aforementioned difficulties, we developed an algorithm as described below. \par

For each pH we have a different objective function,  and  different optimization problems have to be solved, one for each dataset. This results in a set of optimal design parameters for each value of pH, except for $\tilde{B_1}$ that is the ratio of the initial concentration of drug  in the shell to that in the core. Due to the uncertainties on the range $\xi$, in a first stage this feature will not be considered, and the optimization problems are solved independently for each pH, allowing a variation of $\tilde{B_1}$ with pH. Once the optimal values for each pH are available, we select a sub-range of $\tilde{B_1}$ where the objective function is satisfactory for all pH, and a unique value of $\tilde{B_1}$ is selected accordingly. The optimization problems are solved again, detecting the optimal values at each pH and with a common value of $\tilde{B_1}$. Finally, a local optimization is performed in order to verify the correctness of the choice of $\tilde{B_1}$.  \par
The first two steps, where a wide range of parameters is investigated, are resolved by using a {\em Direct Search} global optimization algorithm, while the last step is performed by using a local minimizer.
\bigskip \\
\underline{\em Direct search}

As a first step, a sensible range is defined for five different parameters (\ref{param1}) (table \ref{opt0}). The adopted search algorithm is the Parameter Space Investigation (PSI) \cite{Peri2005}: some sample configurations are uniformly distributed into the design variable space and then evaluated. The uniformity of the distribution of samples is very important since, at the beginning, every part of the variable space has the same probability to contain the global optimum. The search is then concentrated in the neighbourhood of the current best configuration. In order to reduce the number of samples preserving the uniformity of the search,  an Uniformly Distributed Sequence (UDS) is adopted \cite{Statnikov} and is applied for the selection of the candidates. This class of distribution is designed in order to produce a sequence of equally-spaced points. The search is executed in parallel, so that the overall computational time is further reduced.

Due to the wide range of parameters, they are uniformly distributed on a logarithmic scale. In this way, their order of magnitude is more easily detected. We use 
a relatively high number (1024) of samples (not changed along the iterations) to avoid the situation where certain basins of attraction are neglected.
Once all the configurations have been computed, the successive area of investigation is represented by the subspace including the five best locations previously detected. The use of more than a single point is suggested because at the initial stage of the search we have a rather crude analysis of the variable space, so that one could be distracted by a promising point (i.e. a local minimizer) that cannot be further improved, discarding the basin of attraction of the global minimum. This procedure is repeated ten times, providing a successive refinement of the feasible area.
In the second step, PSI is repeated with the same configuration as before, but with a fixed averaged value of $\tilde B_1$. 
\bigskip  \\
\underline{\em Local optimization}

A pattern-search algorithm \cite{Kolda} is adopted for the fine tuning of $\tilde B_1$. Here the total number of parameters is four for each value of the pH plus $\tilde B_1$, that remains the same for all pH values considered. 



\newpage

\subsection{Experimental Methods }

\noindent\underline {Synthesis and Characterization of pH-sensitive hollow nanocontainers} \\
The synthesis and the characterization of this drug delivery system has already been reported elsewhere by some of us \cite{ton2}. Specifically, the synthesis of the pH-sensitive NCs consists of three separate steps: the formation of the sacrificial cores, the synthesis of the shell and the removal of the cores to yield the hollow NCs. 
\smallskip

{\bf PMAA cores: } MAA (2.1 g; 24.4 mmol) was dissolved in 200mL of ACN and stirred at $75^{\circ}C$ for 30min under nitrogen atmosphere, then AIBN (0.3 g; 1.8 mmol) was added and the flask content turned from colorless to a milky suspension. Afterward, the temperature was raised to $(95-100)^{\circ}C$ to  start the distillation. Once 20 mL of ACN were distilled out, the reaction was stopped. The final product was obtained after three cycles of centrifugation and resuspension in ACN (5min $\times$ 8000rpm). 
\smallskip

{\bf Shell:} 0.15g of PMAA cores were suspended in 200mL of ACN and stirred at $75^{\circ}C$ for 30min under nitrogen atmosphere. MAA (0.53 g; 6.1 mmol) was then added, and, after 10 min, PEGMA (0.16g; 0.3 mmol; 5mol \% of MAA) and MBA (0.15 g; 0.97 mmol; 16 mol \% of MAA) were added as well. After 30min, AIBN (0.09g; 0.5 mmol; 8mol \% of MAA) was added. After 10min, the temperature was raised to $(95-100)^{\circ}C$ to start the distillation. Once 30 mL of distilled ACN was collected, the reaction was stopped. The final product was obtained after three cycles of centrifugation and resuspension in ACN (5min $\times$ 5000rpm).  
\smallskip

{\bf Core removal:} 250mg of core-shell structures were suspended in 200mL of a mixture of 1:1 (v/v) $EtOH/H_2O$ and stirred at room temperature overnight. The product was purified by three cycles of centrifugation and resuspension (5min $\times$ 5000 rpm). 
\smallskip 

{\bf Instruments:} NC pictures were taken with scanning electron microscopy (SEM) with W (tungsten) filament operating at 25 kV (Fig. A.7). The hydrodynamic diameter and the $\zeta$ potential of the NCs in distilled water were measured with DLS (Malvern Instruments Series, nano-ZS with multipurpose titrator). The concentration of the sample was 0.1mg/mL, and the given results were the average value of 10 measurements, with 20s integration time. FT-IR spectra were recorded with a PerkinElmer Precisely Spectrum 100 spectrometer. Sonication was performed with an ultrasonic bath (Elmasonic S 30H).

\begin{figure}[ht]
\label{fig9}
\centering
\includegraphics[scale=0.7]{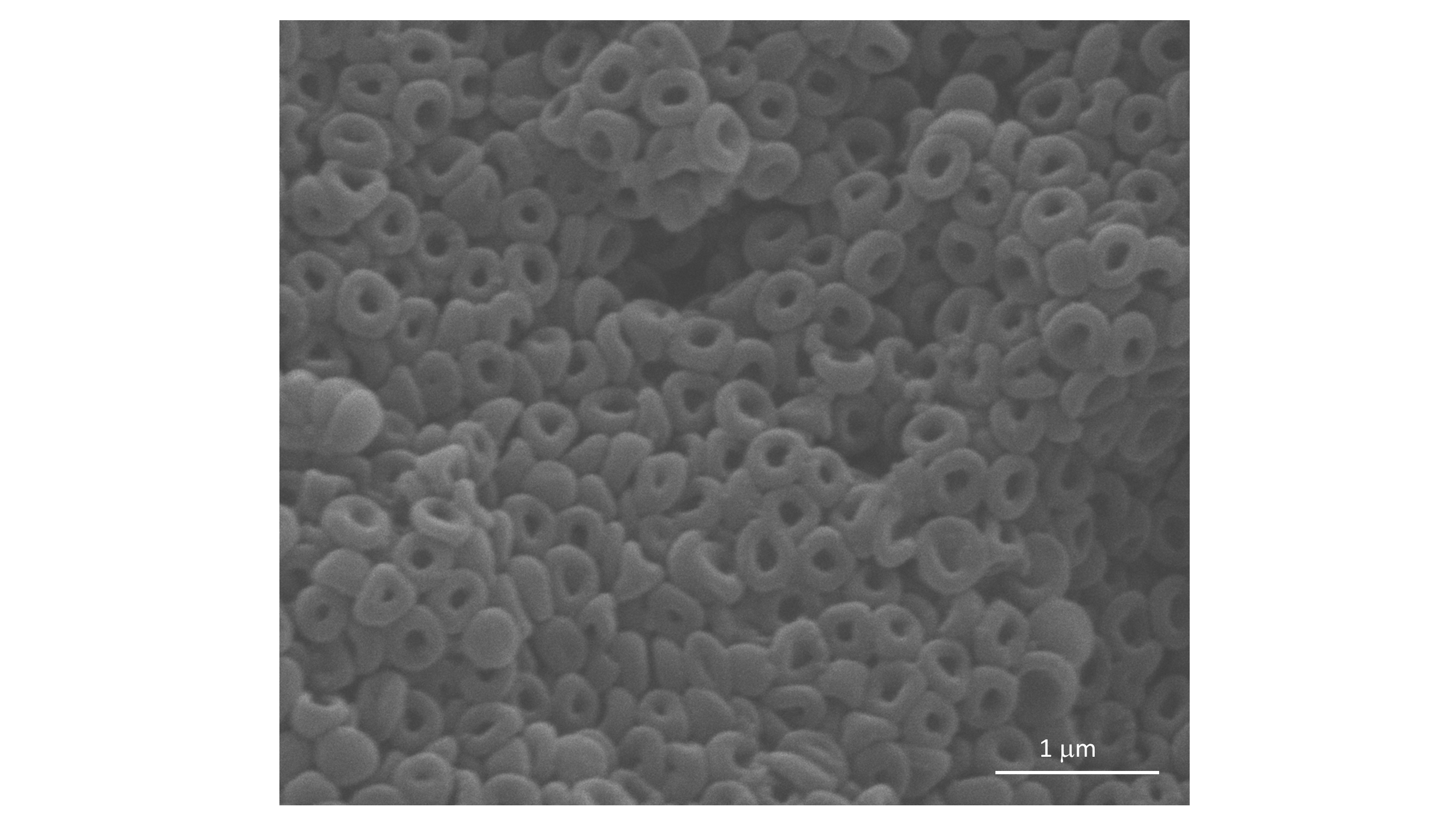} 
\caption{SEM image of pH-sensitive NCs siepmann3}
\end{figure}

\bigskip\bigskip

\noindent\underline{CTP Drug Loading} \\
\noindent 1 mg of NCs was suspended in 950 mL of phosphate-buffered saline (PBS 1$\times$) with the aid of ultrasonic bathing. 2 mg of CTP previously solubilized in 50 mL of ACN were then added to the suspension. The final loading medium, containing 1 mg of NCs and 2 mg of CTP in 1 mL of mixture of PBS/ACN 5\% v/v, was kept under gentle magnetic stirring for 24h at r.t. The non-encapsulated portion of CTP was removed with five cycles of centrifugations and resuspensions in a fresh mixture PBS/ACN (5 min $\times$ 11000 rpm). The amount of encapsulated CTP was indirectly determined by UV spectroscopy and calculated by the difference of concentration between the original CTP solution and the supernatants containing the non-encapsulated drug. The calculations were based on a calibration curve of CTP  obtained in the same solvent mixtures with absorbance measurements at 262 nm \cite{ton2}.

\bigskip\bigskip

\noindent\underline{CTP \textit{in vitro} drug release study} 
\smallskip

1 mg of  CTP-loaded NCs was suspended in distilled water, split into two dialysis bags, and incubated in 25 mL of each release medium: citrate buffer 0.1M + 5\% ACN pH4.0 and PBS 1$\times$ + 5\% ACN pH 7.4. At different time points (as above), 1 mL was collected from each solution and the concentration of the samples was measured using UV spectroscopy. The calculations were made upon a calibration curve of CTP recorded in each buffer (at 262 nm). The experiment was carried out three times for statistical analysis \cite{ton2}.

\end{document}